\documentclass[fleqn,10pt]{SelfArx} 

\usepackage[english]{babel} 

\usepackage{lipsum} 

\definecolor{color1}{RGB}{0,0,90} 
\definecolor{color2}{RGB}{0,20,20} 

\usepackage{hyperref} 
\hypersetup{hidelinks,colorlinks,breaklinks=true,urlcolor=color2,citecolor=color1,linkcolor=color1,bookmarksopen=false,pdftitle={Title},pdfauthor={Author}}

\usepackage{textcomp}

\JournalInfo{\color{white}Journal, Vol. XXI, No. 1, 1-5, 2013} 
\Archive{Additional note} 

\PaperTitle{A Review of Tomographic Reconstruction Techniques for Computed Tomography} 

\Authors{Sanam Assili \textsuperscript{1}*} 
\affiliation{\textsuperscript{1}\textit{Medical Physics Department, School of Medicine, Tabriz University of Medical Sciences, Tabriz, Iran}} 
\affiliation{*\textbf{Corresponding author}: assilis@tbzmed.ac.ir} 

\newcommand{\keywordname}{Keywords} 

\Abstract{Medical imaging modalities have revolutionized health-care approaches by offering a better understanding of the human anatomy. 
Discovery of x-rays allowed the exploiting of the micro-scaled information of human anatomy.
Computed tomography is one of the well-known imaging modalities using x-rays to create images of medical and non-medical objects.
CT imaging has several variety of applications from medical diagnosis to industrial non-destructive testing.
In this paper, we review computed tomography imaging modality and its applications in screening, diagnosis, and treatment and study some of the novel techniques used to reconstruct images in theses machines.}

\begin{document}

\flushbottom 
\maketitle 
\thispagestyle{empty} 


\section*{Introduction} 

\addcontentsline{toc}{section}{Introduction} 

Modern imaging modalities have changed humans’ lives by offering a deeper understanding of the world around us. Discovery of x-rays allowed the exploring of different materials and exploiting the micro-scaled information of human anatomy. X-rays are waves of electromagnetic energy that behave in a similar way as other light rays, but at much shorter wavelengths and can penetrate some thickness of objects. 

X-rays are generated in different ways. The first common process is called Bremsstrahlung approach which is an important phenomenon where x-rays are produced by slowing down the primary beam electrons by the electric field surrounding the nuclei of the atoms in the sample~\cite{low1958bremsstrahlung}. 
Another method is called K-shell emission within which a high-energy electron knocks an electron from an inner orbit in an atom resulting in x-ray emission.
The third method occurs in a synchrotron process, in where a subatomic particle accelerator creates high intensity x-rays that are used for nuclear studies.
The x-ray tubes are used to create the x-ray photons from electric energy supplied by the x-ray generators~\cite{odhner1901nobel}. An x-ray tube is a vacuum tube converting electrical input power into the x-rays. X-ray tubes have evolved from the experimental Crookes tubes that Röntgen implemented his first experiments. Crookes tubes are cold cathode tubes which means that they do not include a heated filament in them to release electrons like the later electronic vacuum tubes. 
A synchrotron is an extremely powerful source of x-rays producing these rays by high energy electrons which circulate around the synchrotron. Synchrotron x-rays can be used for several x-ray imaging approaches, phase-contrast x-ray imaging, and tomography~\cite{winick1997synchrotron,wille1991synchrotron,van1979synchrotron}.
In diagnostic range, x-rays interact with objects via two main processes which are the fundamentals for image formation in a radiographic measurement process. These processes are called the photoelectric effect (photoelectric absorption) and Compton effect or Compton scattering~\cite{sethi2006x,sparks1994resonant}.

In clinical environment, two-dimensional Röntgen images as well as three-dimensional images of human anatomy acquired in computed tomography (CT) scanners are obtained by using x-rays. Several x-ray imaging modalities have been developed based on the different attenuation of x-rays in the structures. Different composition and density of tissues are the causes for the variance in x-ray transmissions. For instance, hard tissues like skeletal structures absorb x-rays more than lowly absorbing parts, such as surrounding tissues.
In this work, we review computed tomography (CT) as one of the well-established x-ray imaging modalities and study different acquisition protocols and image reconstruction approaches used to recover the images in this modality.

\section{Methods}
\label{sec:methods}

In this section, we review computed tomography imaging modality and its applications in screening, diagnosis, and treatment and study some of the novel techniques used to reconstruct images in theses machines. 

\subsection{Computed Tomography}

Computed tomography (CT) is one of the well-known imaging modalities using x-rays to project images. CT imaging has several variety of applications from medical diagnosis to industrial non-destructive testing. 

CT modality has seen different innovations in the past decades which has resulted in an improved modality for diagnosis in clinical indications. The first successful practical implementation of the theory was achieved in 1972 by Sir Godfrey N. Hounsfield, who played a vital role in the development of CT by conducting several experiments based on the mathematical theories of Allan McLeod Cormack in 1964.

\begin{figure}[t]
  \centering
  \includegraphics[width=\linewidth]{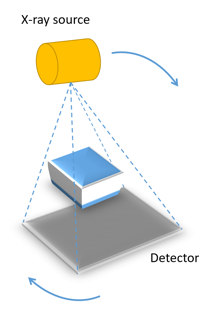}
  \caption{Simple sketch representing the cone-beam technique.}
  \label{fig:fig1}
\end{figure}

A CT scanner combines a series of two-dimensional x-ray projections taken from different angles and uses computer processing to create three-dimensional images, or slices, of a medical sample like bones, blood vessels and soft tissues inside the human body or some industrial materials. Due to the three-dimensional nature of CT scans, this modality provides more detailed information in comparison to a single x-ray image acquisition. In fact, conventional CT scanners are developed to acquire the absorption contrast projections. Due to this reason, CT imaging is one of the mostly used modalities for imaging of hard tissues like bones rather than softer tissues. However, recent advances in x-ray imaging modalities such as contrast enhanced imaging and introduction of phase and dark-field contrast imaging methods have proved a potential for precise measuring and visualizing of softer tissues like hepatic tumors, brain tissue and long nodules. Owing to the recent advancements in mechanics, electronics and computing power, the CT scanning time has been reduced, resultant images have a better quality and readability which helps CT scanners to be chosen as a good non-invasive imaging technology for clinical and non-clinical studies. 

CT modality has several applications in differentiation of soft and hard tissues in head and neck, abdomen and chest such as liver, lung tissue, fat and bone. It is especially useful in detecting for presence, size, spatial location, texture and extent of different types of abnormalities such as lesions, tumors and metastasis within body organs. 
The abdomen contains several organs of the gastrointestinal, urinary, endocrine, and reproductive systems including liver, kidneys, pancreas, spleen, GI tract, and the area around these organs. A CT scan of the abdomen may be performed to assess the abdominal organs for lesions, injuries, or other abnormalities and to investigate the effects of treatments on tumors [Hsi09]. CT scans are frequently performed to evaluate the bones, and joints for damage, lesions, fractures, or other abnormalities, particularly when another type of examination, such as X-rays or physical examination are not conclusive. CT images of the head can provide detailed information about head injuries, severe headaches, dizziness, stroke and brain tumors. These scans are frequently performed to detection of abnormalities and to help diagnosis of unexplained cough, shortness of breath, chest pain, or fever. Lung nodules are detected very commonly on CT scans of the chest, and the ability to detect very small nodules improves with each new generation of CT scanner. 
Industrial CT is an emerging laboratory-based non-destructive testing technique that is used in several applications for inspecting the industrial samples, machine parts and manufactured devices. 

\subsection{Low-dose Computed Tomography Imaging}

Due to radiation risk of using x-rays, research has focused on the radiation exposure distributed to patients while doing CT imaging. 
Therefore, reducing the dose level or low-dose CT has become a significant research area in medical imaging world.
Several methods have been proposed to reduce the dose level by modifying the acquisition protocol~\cite{husmann2007feasibility} or proposing new reconstruction and regularization techniques~\cite{seyyedi2018trpms,seyyedi2017fully3d}.

\subsection{Phase Contrast and Dark-field Imaging}

Phase contrast and dark-field x-ray imaging is an emerging modality that has the potential to improve medical imaging applications and materials analysis. 
Recently, grating interferometer based approaches~\cite{hashimoto2006assessment} have been proposed to extract the scattering and refraction of x-rays by the scanned object in order to obtain phase and dark-field contrast.
Several reconstruction and noise reduction approaches have been proposed to improve the image quality in these methods~\cite{seyyedi2018incorporating,seyyedi2016component}.

\subsection{Artifacts in CT}

Artifacts can seriously degrade the quality of images in computed tomography scans, which could make them diagnostically unusable. To improve image quality, it is essential to understand why artifacts occur and how they can be corrected or removed.

CT artifacts originate due to the range of reasons. Physics-based artifacts occur due to the physical processes in the acquisition process of images. Patient related artifacts are happening due to the several factors associated with patient movement or the presence of metal part in or on the patient body. Scanner related artifacts result from issues in scanner functioning parts. However, in most of the cases, careful patient positioning and precise selection of scanner parameters are the most vital factors to prevent CT artifacts.

\subsection{Tomographic Reconstruction}

Recent advances in computing hardware and subsequently computing power, have opened additional opportunities to improve the performance of CT imaging via more advanced processing approaches, such as tomographic reconstruction methods.
Computed tomography can be assumed as of a mathematical inverse problem, which recovers the attenuation coefficients of a measured object from a set of projection (transmission) values.

If we assume the data collection as a series of parallel rays across a projection at angle $\theta$, the inverse problem in CT can be written based on the Beer-Lambert law, which describes the absorption of x-rays as,
\begin{equation} \label{eqn_res_norm}
I=I_0e^{-\int\mu(x,y)ds},
\end{equation}
where $\mu(x,y)$ denotes the attenuation coefficient which is specific to each object type and $I$ and $I_0$ refer to the transmitted and incident intensities respectively.

\subsubsection{Analytical Reconstruction Methods}

Analytical methods are a commonly used category of image reconstruction techniques for CT imaging modality.
The most commonly used analytical reconstruction methods on commercial CT scanners are all in the form of filtered back-projection (FBP), which uses a one-dimensional filter on the projection data before back-projecting (two- or three-dimensional) the data onto the image space~\cite{geyer2015state,oliveira2011comparison,feldkamp1984practical}.

Several limitations reduce the performance of analytical reconstruction methods for practical medical and industrial scenarios. 
These methods generally ignore associated noise of measurements in the problem and tries to reduce this artifact's effect by post-filtering operations.

\subsubsection{Iterative Reconstruction Techniques}

\begin{figure}[!h]
  \centering
  \includegraphics[width=\linewidth]{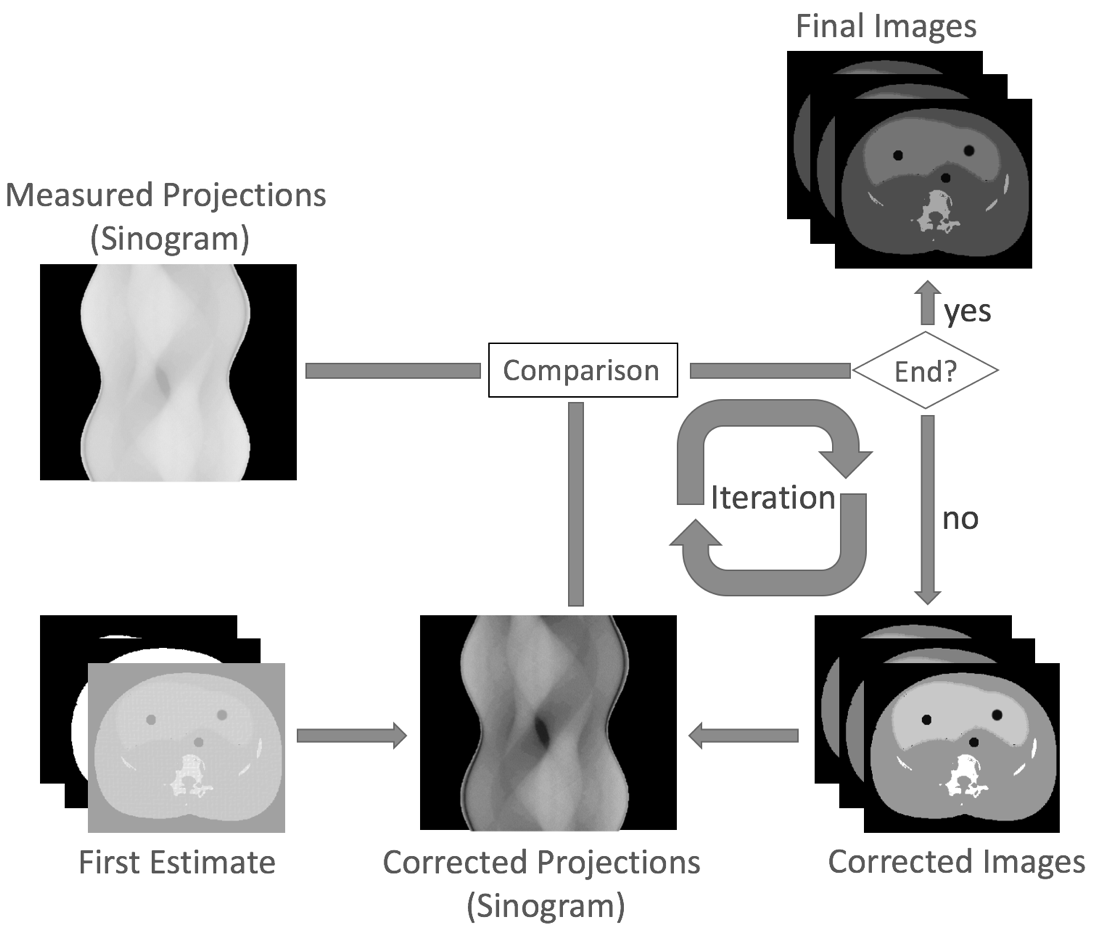}
\caption{A simplified schematic for iterative reconstruction techniques for CT modality.}
\label{fig:ir_schematic}
\end{figure}

Iterative reconstruction refers to a category of algorithms used in CT imaging that begin with an image assumption, and compares it to the real time measured values while making constant adjustments until these two are in agreement`\cite{herman2009fundamentals,hu1999multi}.
The principle of iterative image algorithms is illustrated in~\autoref{fig:ir_schematic} in several steps. 

As shown in this figure, following a CT acquisition process to measure projections, a first image estimation is generated. 
An x-ray beam is simulated via forward projection to obtain simulated projection data, which are then compared with the measured projection data. 
In case of difference, the first image estimation will be updated based on the features of the underlying method~\cite{sey2014embc}.

\paragraph{Algebraic Reconstruction Technique (ART)} 

The algebraic reconstruction technique (ART) was the first widely used iterative approach with a long history and rich literature. 
It was first introduced by Kaczmarz in 1937~\cite{kaczmarz1937angenaherte} and was independently used by Gordon et al.~\cite{gordon1970algebraic} in image reconstruction. 
ART is a reconstruction algorithm that uses a set of projections to reconstruct the desired object.

Assuming the original linear problem $AX = Y$, we can write,

\begin{equation} \label{eqn:art_system}
\sum_{j=1}^Na_{ij}x_{j}=y_{i}, \quad s.t. \quad i=1,2,...,M \quad and \quad j=1,2,...,N, 
\end{equation}
where $a_{ij}$ is the weighting parameter which denotes the influence of $i$th cell on the $j$th line integral, $x_j$ is the constant intensity value of the $j$th cell, $N$ refers to the total number of cells, and $M$ refers to the total number of rays~\cite{seyyedi2013object,sey3013ispa,seyyediehbpaper}. 

\paragraph{Simultaneous Algebraic Reconstruction Technique (SART)}

In 1984, the simultaneous algebraic reconstruction technique (SART) was introduced with major changes in the standard ART approach.
SART had a major impact in CT imaging scenarios with limited projection data. 
It generates a good reconstruction in just one iteration and illustrates superior performance comparing to the original ART approach~\cite{avinash1988principles, seyyedi2014sayisal}.

\paragraph{Maximum Likelihood Expectation Maximization (MLEM)} 

The methods introduced so far are assuming well-posed problem with some good measurements and none of them model the statistical properties of the measurement process.

Likelihood based approaches are another category of methods for photon-limited conditions that are the standard since decades, in order to support low dose imaging~\cite{vardi1985statistical,lange1984reconstruction}.

\paragraph{Penalized Likelihood (PL)} 

Penalized likelihood (PL) estimation is a way to consider the complexity of a model while estimating parameters of different models. 
In general, instead of applying a simple MLE, the log-likelihood minus a penalty term will be maximized, which is depending on the model and most often increasing with number of parameters~\cite{fessler2000statistical}.

\section{Conclusion}

In conclusion, CT imaging has come a long way in last decades since Rontgen discovered x-rays and Hounsfield designed the first clinical CT machine. 
Since then, several novel x-ray and CT based imaging modalities have been proposed to visualize more advanced details in the scanned targets. Among them, are phase contrast and dark-field imaging techniques that have the potential to improve medical imaging applications and materials analysis.
Owing to the recent advances in computing hardware and subsequently computing power, improvements in the performance of CT imaging via more advanced processing approaches, such as tomographic reconstruction methods have been developed.
Novel dose reduction techniques have been utilized to extend the wide spread use of CT imaging for medical scenarios. Therefore, computed tomography imaging modality has become an important technique in screening, diagnosis, and treatment and study several diseases and abnormalities.

\section*{Disclosure statement} 

\addcontentsline{toc}{section}{Disclosure statement} 

The author declares there is no conflict of interest regarding the publication of this manuscript.

\phantomsection
\bibliographystyle{unsrt}


\end{document}